\documentclass[aip,apl,reprint,dvipdfmx]{revtex4-1}

\usepackage{graphicx}
\usepackage{amsmath}
\usepackage{dcolumn}
\usepackage[bookmarks=false,colorlinks=true,linkcolor=blue,citecolor=blue,urlcolor=blue]{hyperref}
\usepackage{xr}
\usepackage{here}

\begin{document}

\title{Possible electronic entropy-driven mechanism for non-thermal ablation of metals}

\author{Yuta Tanaka}
\email[Electronic mail: ]{tanaka@cms.phys.s.u-tokyo.ac.jp}
\affiliation{Department of Physics, The University of Tokyo, 7-3-1 Hongo, Bunkyo-ku, Tokyo 113-0033, Japan}
\author{Shinji Tsuneyuki}
\affiliation{Department of Physics, The University of Tokyo, 7-3-1 Hongo, Bunkyo-ku, Tokyo 113-0033, Japan}
\affiliation{Institute for Solid State Physics, The University of Tokyo, 5-1-5 Kashiwanoha, Kashiwa, Chiba 277-8581, Japan}

\date{\today}

\begin{abstract}
The physical mechanism for metal ablation induced by femtosecond laser irradiation was investigated. 
  Results of calculations based on finite-temperature density functional theory (FTDFT) indicate that condensed copper becomes unstable at high electron temperatures due to an increase of electronic entropy at large volume, where the local density of states near the Fermi energy increases. 
  Based on these results, an electronic entropy-driven (EED) model is proposed to explain metal ablation with a femtosecond laser.
  In addition, a mathematical model is developed for simulation of the laser ablation, where the effect of the electronic entropy is included.
 This mathematical model can quantitatively describe the experimental data 
  in the low-laser-fluence region, where the electronic entropy effect is determined to be especially important.

 \end{abstract}


\maketitle

Specific phenomena have been observed by irradiation of a metal surface with a femtosecond laser, such as ultrafast structural changes,~\cite{Fritz_2007,Daraszewicz_2013} bond hardening,~\cite{Ernstorfer_2009} and the emission of excessively high energy ions and neutral atoms.~\cite{Miyasaka_2012,Dachraoui_2006,Dachraoui_2006_2}
The process of  removing materials with an intense laser is called ablation, and ablation which cannot be explained under the assumption of thermal equilibrium  such as the last one is referred to as non-thermal ablation.
One of the reasons for the specific attention to non-thermal ablation from industry is that it can decrease the thermal damage region.~\cite{Chichkov_1996,Shaheen_2013,Momma_1996}
However, a complete understanding of non-thermal ablation of metals is still missing, so that optimal laser conditions for precision processing with femtosecond laser irradiation cannot be predicted from simulation.

Femtosecond laser irradiation on a metal surface changes the electron subsystem of the metal from the ground state into excited states.
An electron subsystem is thermalized to the Fermi-Dirac distribution with  electron temperature $T_e$, via electron-electron interaction, of which the scattering time $\tau_{ee}$ is approximately $10 \mathchar`-100\,\text{fs}$ in metals.~\cite{Mueller_2013, Brown_2016_2}
At the same time, the lattice  temperature $T_l$ begins to increase by energy transfer from the electron subsystem via electron-phonon interaction, of which the scattering time $\tau_{el}$ is on the time scale of picoseconds.~\cite{Schoenlein_1987,Elsayed-Ali_1987,Elsayed-Ali_1991,Hohlfeld_2000}
Therefore, under the assumption of instantaneous and local thermalization in the electron subsystem,
$T_e \gg T_l$ is expected to be realized long before $\tau_{el}$ by intense femtosecond laser irradiation.
This is the main concept of the two-temperature model (TTM).~\cite{Anishimov_1973}

The TTM has been widely employed to simulate~\cite{Daraszewicz_2013,Ernstorfer_2009,Giret_2011,Norman_2012,Norman_2013,Recoules_2006} such phenomena induced by femtosecond laser irradiation and
 it has been successful for quantitative description of the experimental data.~\cite{Daraszewicz_2013,Ernstorfer_2009}
These TTM calculation results indicated that change of the interatomic forces due to high $T_e$ is required to reproduce the experimental data.
Recently, some calculation studies have suggested that ablation can be caused by a change of the electronic states without emitting electrons from a metal surface.~\cite{Norman_2012,Norman_2013}
This explanation, which does not require the neutrality breakdown, is supported experimentally,~\cite{Li_2011}
although there is a conflict with a previous explanation based on the Coulomb explosion (CE) model,~\cite{Dachraoui_2006, Dachraoui_2006_2, Miyasaka_2012} which is verified in an insulator~\cite{Stoian_2000_2} and a molecule.~\cite{Sato_2008}
Therefore,  in case of metals, where many  electrons with good mobility exist, we expect that the explanation based on the CE model should be replaced by the former explanation.
Although these previous studies have suggested that a large contribution of the force induced by change of the electronic states originates from the kinetic energy of free (delocalized) electrons,~\cite{Norman_2012,Norman_2013,Stegailov_2015,Stegailov_2016} the validity of this explanation has yet to be clarified.
Furthermore, comparison with experimental results has not been performed sufficiently.

Here, we first investigate the physical mechanism of non-thermal ablation of copper (Cu) based on results of first-principles calculations, and propose the electronic entropy-driven (EED) model to describe it.
Subsequently, a mathematical model is developed for simulation of the ablation depth, where the EED model is employed.
Finally, we present results of the simulation and compare them with the experimental data.~\cite{Colombier_2005, Nielsen_2010}  
It should be noted that the TTM is employed in all our calculations.

To investigate the main contribution that causes non-thermal ablation, we conducted first-principles calculations based on finite-temperature density functional theory (FTDFT).~\cite{Mermin_1965}
Before thermal equilibrium is achieved, a system irradiated with an intense femtosecond laser can be represented as $T_e > T_l$ by employing the TTM.
To examine $T_e$ dependence of the stability of Cu, the electronic free energy $F$ was calculated, which is defined as:
\begin{equation}
    F =  E -T_e S,  \label{eq: Free_energy} 
\end{equation}
where $E$ is the internal energy and $S$ is the electronic entropy.
$S$ for independent particles that occupy single-particle states is written as:
\begin{equation}
    S = - 2 k_B \sum_i [f(\epsilon _i)\ln f(\epsilon _i)+ (1- f (\epsilon _i)) \ln (1- f (\epsilon _i))],  \label{eq: Entropy} 
\end{equation}
where $f (\epsilon _i)$ is the occupation of the eigenenergy $\epsilon _i $, where  $i$ denotes an eigenstate, the sum is over one-electronic eigenstates, and $k_B$ is the Boltzmann constant. 
In thermal equilibrium state with respect to an electron subsystem, the occupation $f(\epsilon _i)$ can be expressed as the Fermi-Dirac distribution, $f(\epsilon _i) = (1 +  e^{(\epsilon _i - \mu)/k_BT_e})^{-1}$, where $\mu$ is the chemical potential.
To simplify the calculations, only the volume dependence at each $T_e$ was considered.
The face centered cubic (fcc) structure for primitive cells of Cu at different volumes are calculated in a range of $T_e$ between $300$ and $25000\,\text{K}$. 
To analyze the volume dependence of $S$ at high $T_e$, we calculated the band structure and the density of states (DOS) at $V_0$, which is the equilibrium volume  at $T_e = 300\,\text{K}$, and at $2V_0$.
Both calculations were conducted at $T_e = 25000\,\text{K}$. 
The relative error between our calculation value $V_0 = 12.23\,$\AA$^3$ and an experimental~\cite{Straumanis_1969} value $V_{\text{exp}} = 11.81\,$\AA$^3$ is $3.6\,\%$.

%%%%%%% Fig 1 (tanaka) %%%%
\begin{figure}[t]
  \includegraphics{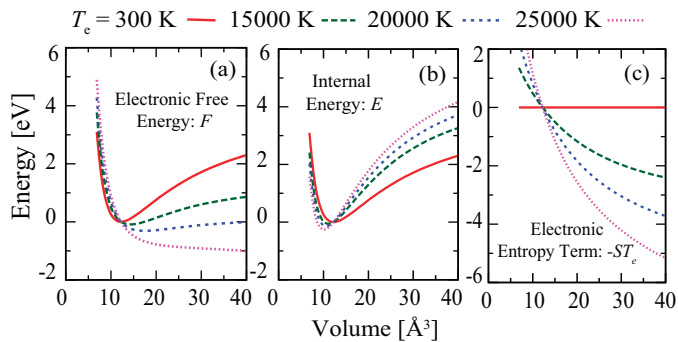}
  \caption{
    \label{fig: FES}
     $T_e$ dependence of (a) the electronic free energy $F$, (b) the internal energy $E$, and (c) the electronic entropy term $-ST_e$, as a function of the fcc primitive cell volume.
    The bases of the vertical axes are set to each value at $V_0$.
    }
\end{figure}
%%%%%%%%%%%%%%%%%%%%

These calculations were performed using xTAPP code,~\cite{xTAPP} in which the electronic entropy $S$ calculation was implemented.
The ultra-soft pseudopotential and the generalized gradient approximation (GGA) with the Perdew--Burke--Ernzerhof exchange-correlation functional~\cite{Perdew_1996,Perdew_1997} were used.
In the ultra-soft pseudopotential, the $3d^{10}$ and $4s^1$ states are treated explicitly as valence states.
The electronic structures were calculated with a cutoff energy of $1200\,\text{eV}$ for the plane-wave basis and the Brillouin-zone $k$-point sampling of a Monkhorst-Pack mesh with $12\times12\times12$ $k$-points for the fcc primitive cell.
The number of bands was 13.
In the DOS calculation, only a number of the Brillouin-zone $k$-point sampling was altered to $16\times16\times16$ $k$-points.

%%%%%%% Fig 2 (tanaka) %%%%
\begin{figure}[b]
  \includegraphics{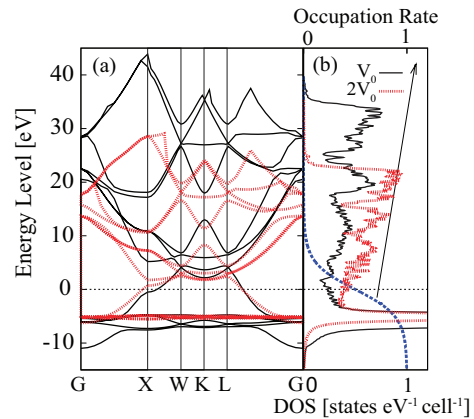}
  \caption{
    \label{fig: BAND_DOS}
    (a) Band structures and (b) DOS at $T_e = 25000\,\text{K}$. 
    Solid black and red dotted lines represent $V_0$ and $2V_0$, respectively.
    The blue dashed line is the Fermi-Dirac distribution of $T_e = 25000\,\text{K}$.
  }
\end{figure}
%%%%%%%%%%%%%%%%%%%%

The calculation results for $F$, $E$ and $-ST_e$ as a function of $V$ at $T_e = 300, 15000, 20000$ and $25000\,\text{K} $ are shown in Fig.~\ref{fig: FES}. 
 Fig.~{\hyperref[fig: FES]{\ref*{fig: FES}(a)}} shows that the curvatures of electronic-free-energy curves become smaller as the values of $T_e$ increase.
Between $300$ and $20000\,\text{K}$, the minimum points that correspond to equilibrium volume at each $T_e$ shift to larger values.
Finally, between $20000$ and $25000\,\text{K}$, the minimum point vanishes. 
These results are qualitatively consistent with previous studies for tungsten.~\cite{Murphy_2015,Khakshouri_2008} 
The present results indicate that if atoms can freely change their interatomic distance, such as atoms near a surface, then they cannot be condensed around $ T_e = 25000\,\text{K}$ under the assumption of an isothermal process with respect to $T_e$.
The validity of this assumption will be discussed later.

To discuss the main contribution for the disappearance of the minimum point in Fig.~{\hyperref[fig: FES]{\ref*{fig: FES}(a)}}, $E$ and  $-ST_e$, which are components of $F$, are plotted in Figs.~{\hyperref[fig: FES]{\ref*{fig: FES}(b)}} and {\hyperref[fig: FES]{(c)}}, respectively.
Fig.~{\hyperref[fig: FES]{\ref*{fig: FES}(b)}} shows that the values of $E$ at high $T_e$ are larger than these at low $T_e$ in the region of $V > V_0$.
On the other hand, Fig.~{\hyperref[fig: FES]{\ref*{fig: FES}(c)}} shows that the values of $-ST_e$ at high $T_e$ are smaller than these at low $T_e$ in the region of $V > V_0$.
Hence, we find that the main contribution for the disappearance of the electronic free energy minimum originates from $-ST_e$.
To discuss the reason for this large benefit of $S$ at large volume, the band structures and the DOS at different volumes of $V_0$ and  $2 V_0$ are plotted in Fig.~\ref{fig: BAND_DOS}.
From Fig.~{\hyperref[fig: BAND_DOS]{\ref*{fig: BAND_DOS}(b)} and the definition of $S$ (Eq.~(\ref{eq: Entropy})), the local DOS  near the Fermi energy at a large volume of  $ 2 V_0$ is larger than those at $V_0$ and this change increases the value of $S$.
This behavior can be easily understood as a decrease of the hopping energy between atoms due to the increased interatomic distance at the large volume. 
These consideration for the benefit of $S$ were also suggested in a previous study.~\cite{Khakshouri_2008} 

We conclude that atoms cannot be condensed around $25000\,\text{K}$ due to an increase of the electronic entropy $S$, and consequently non-thermal ablation occurs.
We call this physical mechanism the EED model.
It should be noted that this explanation does not require the neutrality breakdown, which is denied by the experimental result~\cite{Li_2011}  in the case of metal ablation.
Based on these results, we propose the EED model to explain the non-thermal ablation of metals, not only for copper, because the physical explanation given here is expected to be applicable to all metals.

Subsequently, a mathematical model that included the EED model was developed for simulation of the ablation depth, and results of Cu films calculated by employing this model are presented.
One of purposes of these calculations is to validate the mathematical model and the EED model.
The other is to discuss the contribution of $S$ to the ablation depth.
$T_e$ and $T_l$  were calculated by solving the following two-coupled differential equations for the electron (Eq.~(\ref{eq:electron})) and the lattice (Eq.~(\ref{eq:lattice})) subsystems,~\cite{Anishimov_1973}
\begin{subequations}
\begin{align}
 C_e \frac{{\partial}T_e}{{\partial}t}  &= {\nabla }\cdot(\kappa _e{\nabla }T_e) - G(T_e - T_l) + S_{\text{laser}},  \label{eq:electron} \\   
 C_l \frac{{\partial}T_l}{{\partial}t} &= {\nabla}\cdot(\kappa _l{\nabla}T_l) + G(T_e - T_l).      \label{eq:lattice}
\end{align}
\end{subequations}
Here, $C$ and $\kappa $ are the heat capacity and the thermal conductivity, respectively.
The $e$ and $l$ indices denote the electron and lattice subsystems, respectively. 
$G$ is the electron-phonon heat transfer constant and $S_{\text{laser}}$ is a source term that describes the energy deposition by the laser pulse.
The thermal diffusion along a surface can be neglected because the sum of the laser penetration depth $\delta = 13\,\text{nm}$~\cite{Johnson_1972} and the mean free path of electrons $\delta_{\text{mfp}} = 42\,\text{nm}$~\cite{Ashcroft_1976} of Cu is much smaller than the radius of the irradiated laser spot.
Accordingly, the three-dimensional Eqs.~(\ref{eq:electron}) and (\ref{eq:lattice}) can be reduced to one-dimensional equations.
In addition, the melting of Cu is neglected in our calculations.
This assumption is expected to be suitable in the case of the low-fluence laser irradiation because molten materials were not detected experimentally~\cite{Chichkov_1996,Shaheen_2013,Momma_1996} for these laser condition.

To accurately calculate time and space evolution of $T_e$ and $T_l$, close attention should be paid to determine these parameters in Eqs.~(\ref{eq:electron}) and (\ref{eq:lattice}).
$T_e$ dependent heat capacity $C_e(T_e)$ is obtained by fitting previous calculation results,~\cite{Bvillon_2015, Lin_2008} where $C_e(T_e)$ is calculated by taking the derivative of the internal energy $E(T_e)$ with respect to $T_e$. 
According to the Dulong-Petit law, $C_l = 3.51\,\text{J}\,\text{cm}^{-3} $ is given by $V_{\text{exp}}$.
$C_l$ can be assumed to be constant above the Debye temperature $T_D = 343\,\text{K}$~\cite{Papaconstantopoulos_1977} so that this value is a good approximation in our simulation, where $T_l > T_D$ is almost always satisfied.
Based on the Drude model, $\kappa_e(T_e,T_l) = \frac{1}{3} v_F^2 C_e(T_e) \tau _e(T_e,T_l)$ can be derived.
Here, $v_F = 1.57 \times 10^6\,\text{m}/\text{s}$~\cite{Ashcroft_1976} and $\tau_e(T_e,T_l)$ are the Fermi velocity and the electron relaxation time, respectively.
According to the Fermi liquid theory, $\tau_e(T_e,T_l)$ for electrons with energy near the Fermi energy is approximated as:
$ \tau _e^{-1}(T_e,T_l) =  \tau _{ee}^{-1}(T_e) +  \tau _{el}^{-1}(T_l)  = A_e T_e ^2 + B_l T_l$,
where $A_e$ and $B_l$ are typically assumed to be constant.~\cite{Wang_1994,Kaveh_1984}
$B_l = 1.98\times10^{11}\,\text{s}^{-1} \text{ K}^{-1}$ was determined so as to reproduce experimental value~\cite{Laubitz_1967} $\kappa _e = 3.99\, \text{ W} \text{ cm}^{-1}  \text{ K} ^{-1}$ at low temperature.
$A_e=2.22 \times 10^6 \,\text{ s}^{-1} \text{ K}^{-2}$  was obtained  by a recent  first-principles calculation.~\cite{Brown_2016}
This value is consistent with the experimental result,~\cite{Kaveh_1984} in which $6.68 \times 10^5 < A_e < 2.89\times10^6\,\text{ s}^{-1}\text{ K}^{-2} $ was reported.
$\kappa_l$ is often neglected for pure metals because this value is much smaller than that of $\kappa_e$. 
$G =1.0\times10^{17}\,\text{ W} \text{ K}^{-1} \text{m}^{-3}$ was obtained by first-principles calculation,~\cite{Migdal_2016}
where it was suggested that the $T_e$ dependence of $G$ is small at least below $32000\,\text{K}$. 
In our calculation, this condition is satisfied for a laser peak fluence $F <1.5\,\text{J} \text{ cm}^{-2}$.
The source term $S_{\text{laser}}$ is assumed to be:~\cite{Hohlfeld_2000}  
\begin{equation}
   \begin{split}
S_{\text{laser}}  = \sqrt{ \frac{ \beta }{\pi }} \frac{(1-R)F}{t_p(\delta + \delta _b) }  
              \exp \Bigg[ - \frac{z}{\delta  + \delta _b} - \beta \Big(\frac{t - t_0 }{t_p} \Big)^2  \Bigg],  \label{eq:S}
    \end{split}
\end {equation}
where
$z$ is a depth spacial coordinate, 
$t$ is the elapsed time, 
$R$ is the reflectivity,
$t_p$ is the laser pulse duration time,
$t_0$ is the delay time of the laser,
$\delta _b$ is the ballistic range of electrons,
and $\beta = 4 \ln 2 $.
$R$ was recently reported to depend on the number of irradiated pulses $n$ and laser fluence $F$ because of the laser-structured surface and the change of the dielectric constant of the irradiated material.~\cite{Vorobyev_2011}
Therefore, to compare the simulation results with the experimental data,~\cite{Colombier_2005,Nielsen_2010} where over approximately a few tens~\cite{Colombier_2005} or one hundred~\cite{Nielsen_2010} laser pulses are irradiated, 
the change of reflectivity must be considered.
However, it is too difficult to consider these effects without experimental data.
Therefore, in our calculations,  $R_n(F)$ were determined by fitting the experimental reflectivity~\cite{Vorobyev_2011} as $R_n(F) = a_n\ln F + b_n$, and  simulations were conducted for each reflectivity $R_n(F)$ ($n =10, 50, 100$).
As results of $a_n$ and  $b_n$ calculations, good $R_n$ were obtained, of which the root mean square errors were less than 0.02.
The laser penetration depth $ \delta(T_e) $ was determined by a critical point model~\cite{Etchegoin_2006,Etchegoin_2007} and parameters were obtained from a previous study.~\cite{Ren_2011} 
The ballistic range was approximated as:~\cite{Hohlfeld_2000} $ \delta _b(T_e,T_l) =\tau _e(T_e,T_l) v_F$.

To solve Eqs.~(\ref{eq:electron}) and (\ref{eq:lattice}), the finite-difference methods were used, where time $\Delta t$, and space step $\Delta z$, were $10\,\text{as}$ and $1\,\text{nm}$, respectively.
The Neumann boundary condition was used.
The thickness of the calculated Cu film was $1\,{\mu}\text{m}$.
The Cu film was irradiated on the front surface by a laser with the laser pulse duration time $t_p$ of $100\,\text{fs}$ and a wavelength of $800\,\text{nm}$.
The delay time $t_{0}$ was $4t_p$ and both the initial $T_{e}$ and $T_{l}$ were set to $300\,\text{K}$.
 It should be noted that the fitting parameters were not used in the simulation.

%%%%%%% Fig 3 (tanaka) %%%%
\begin{figure}[t]
  \includegraphics{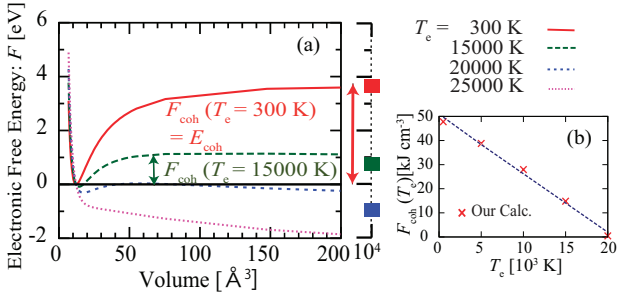}
  \caption{
    \label{fig: E_stable}
    (a) $T_e$ dependence of $F$ as a function of the primitive fcc structure volume. 
    (b) $T_e$ dependence of $F_{\text{coh}}(T_e)$. Cross marks represent the calculated data and the dotted line represent the fitting of these data as a linear function.
  }
\end{figure}
%%%%%%%%%%%%%%%%%%%%%%%%%

%%%%%%% Fig4 (tanaka) %%%%
\begin{figure}[b]
  \includegraphics{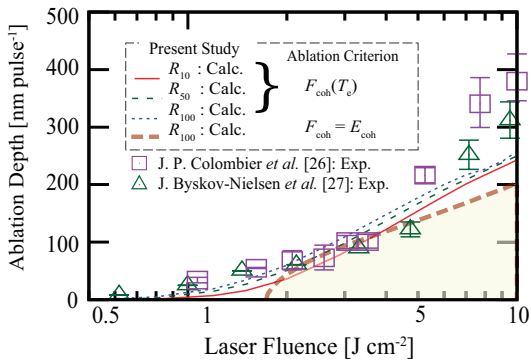}
  \caption{
    \label{fig: Depth}
    Comparison of ablation depth from the simulation results and the experimental data. 
    $R_n$ denotes the reflectivity of the n-th laser irradiation pulse was used in these simulations. 
    Thin lines and the bold line represent our calculation results, where  $F_{\text{coh}}(T_e)$ and a constant $F_{\text{coh}} = E_{\text{coh}}$ were used as the criteria for ablation (Eq.~(\ref{eq:criteria})), respectively.
    Square~\cite{Colombier_2005} and triangle~\cite{Nielsen_2010} symbols represent the experimental data.
     The ablation threshold laser-fluences at each $R_{n}(F)$ ($n=10$, $50$, $100$) are  $0.77$, $0.58$, and $0.52\,\text{J\,cm}^{-2}$, respectively. 
      }
\end{figure}
%%%%%%%%%%%%%%%%%%%%

It was assumed that to cause ablation, the lattice energy $E_l (T_l(t,z))$ at a grid of $z$, which is defined as $E_l (T_l(t,z))= C_l \,T_l(t,z)$, must overcome the largest values of electronic free energy $F_{\text{coh}}(T_e(t,z))$ (Fig.~\ref{fig: E_stable}) in the region of $V>V_0$.
At low $T_e$, $F_{\text{coh}}(T_e)$ corresponds to the cohesive energy $E_{\text{coh}} = 47.76\,\text{kJ}\,\text{cm}^{-3}$, the value of which is the result of our calculation (Fig.~{\hyperref[fig: E_stable]{\ref*{fig: E_stable}(b)}}), and which agrees well with the experimental value~\cite{Straumanis_1969,Kittle} $47.34\,\text{kJ}\,\text{cm}^{-3}$.
In addition to this assumption, we assume that ablation occurs only at a grid of the surface $z_{\text{surf}}$ because the bulk cannot expand freely.
  Taken together, the criterion for ablation can be expressed as the following inequality:
\begin{equation}
    E_l(T_l(t,z_{\text{surf}})) > F_{\text{coh}} (T_e(t,z_{\text{surf}})).  \label{eq:criteria}
\end{equation}
Fig.~{\hyperref[fig: FES]{\ref*{fig: FES}(b)}} shows that the values of the internal energy $E$ at large volume are larger than those of $E$ at $V_0$, even at high $T_e$. 
Therefore, the absorption of the latent heat $E_{\text{late}}$ is required for ablation.
In the present simulation, $E_{\text{late}}$ is assumed as:
\begin{equation}
    E_{\text{late}}(T_l (t,z_{\text{surf}})) = E_{\text{coh}}  - E_l(T_l(t,z_{\text{surf}})).  \label{eq:latent}
\end{equation}
If the grid of $z_{\text{surf}}$ satisfies Eq.~(\ref{eq:criteria}), then $E_{\text{late}}(T_l)$ begins to be absorbed as the latent heat from the electronic subsystem at the grid of $z_{\text{surf}}$.
We consider that the delay time $t_{\text{abs}} = \Delta z/{v_{\text{s}}}$ is required to cause ablation after a grid becomes the grid of $z_{\text{surf}}$  because ablation wouldn't occur until passing through a pressure wave, which is created by previous ablation.
The velocity of the pressure wave is assumed to be the velocity of sound, $v_{\text{s}}=4760\,\text{m}\,\text{s}^{-1}$.~\cite{Linde_2003}
To represent ablation, the grid of $z_{\text{surf}}$ is removed from the simulation, and the  grid of $z_{\text{surf}}+1$ becomes the new surface grid after the absorption of $E_{\text{late}}(T_l)$.

The calculation results and experimental data~\cite{Colombier_2005, Nielsen_2010} are plotted in Fig.~\ref{fig: Depth}.
The thin lines indicate that the dependence on the number of pulses is not large between the $10$ and $100$th pulses.
Moreover, these thin lines indicate that the calculation results with consideration of the electronic entropy $S$ effects are in good agreement with the experimental data~\cite{Colombier_2005, Nielsen_2010} in the low-laser-fluence region ($\lesssim5\,\text{J}\,\text{cm}^{-2}$), where the effect of non-thermal ablation is expected to be dominant.~\cite{Chichkov_1996,Shaheen_2013,Momma_1996}
However, the gradients of these lines in the high-laser-fluence region ($> 5\,\text{J}\,\text{cm}^{-2}$) are underestimated.
We consider that the disagreement in the high-laser-fluence region is due to a lack of physical mechanics, such as the ejection of liquid droplets by the recoil pressure~\cite{Chichkov_1996} created by ablation.
On the other hand, in the low fluence region, this effect is expected to be little because molten material isn't created~\cite{Chichkov_1996,Shaheen_2013,Momma_1996} in this region.

The dashed bold line in Fig.~\ref{fig: Depth} represents calculation results when only thermal ablation is considered, in which the effect of $S$ is ignored.
In other words, in these calculations, constant cohesive energy $F_{\text{coh}} = E_{\text{coh}} $ was used instead of $F_{\text{coh}}(T_e)$ for the ablation criterion in Eq.~(\ref{eq:criteria}).
This line has no tail in the low-laser-fluence region; therefore, we suggest that non-thermal ablation in the low-laser-fluence region is caused by the electronic entropy effects.
It should be noted that in our simulation, the change of $T_e$ at the grid of $z_{\text{surf}}$ during $t_{\text{abs}}$ is approximately $10\mathchar`-15\%$ due to thermal flux from deeper grids.
Therefore, the isothermal process with respect to $T_e$ is expected to be conserved at the grid of $z_{\text{surf}}$ during $t_{\text{abs}}$.

In summary, the results of FTDFT calculations show the instability of condensed Cu at high $T_e$ due to the electronic entropy effect.
Furthermore, the results of the band structure and the DOS indicate that the volume-dependence in the electron states near the Fermi energy is the main contribution to this instability.
Based on these results, we propose the EED model to describe the physical mechanism for the non-thermal ablation of metals.
The results of the developed mathematical model calculations, where the effect of the electronic entropy $S$ is included, show that this model can predict the experimental data in the low-laser-fluence region. This strongly suggests that the electronic entropy effect is dominant for non-thermal ablation of metals.

\begin{acknowledgments}
  This work was supported in part by the Innovative Center for Coherent Photon Technology (ICCPT) in Japan.
  Y. T. was supported by the Japan Society for the Promotion of Science through the Program for Leading Graduate Schools (MERIT)\@.
 \end{acknowledgments}

\end{document}